\documentclass[twocolumn,prb,amsmath,showpacs,amssymb,citeautoscript,floatfix]{revtex4}
\usepackage{graphicx,color}

\usepackage{dcolumn}
\usepackage{bm,ulem}
\usepackage{amsmath,amssymb}
\begin{document}

\title{Defect-induced magnetic structure of CuMnSb}

\author{F. M\'aca}
\affiliation{Institute of Physics ASCR, Na Slovance 2, CZ-182 21 Praha 8, Czech Republic}

\author{J. Kudrnovsk\'y}
\affiliation{Institute of Physics ASCR, Na Slovance 2, CZ-182 21 Praha 8, Czech Republic}

\author{V. Drchal}
\affiliation{Institute of Physics ASCR, Na Slovance 2, CZ-182 21 Praha 8, Czech Republic}

\author{I. Turek}
\affiliation{Charles University, Faculty of Mathematics and
Physics, Department of Condensed Matter Physics, Ke Karlovu 5,
CZ-121 16 Prague 2, Czech Republic}

\author{O. Stelmakhovych}
\affiliation{Charles University, Faculty of Mathematics and
Physics, Department of Condensed Matter Physics, Ke Karlovu 5,
CZ-121 16 Prague 2, Czech Republic}

\author{P. Beran}
\affiliation{Nuclear Physics Institute ASCR, CZ-250 68 \v{R}e\v{z}, Czech Republic}

\author{A. Llobet}
\affiliation{Neutron Science and Technology, Physics,
 Los Alamos National Laboratory, Los Alamos, NM 87544, USA}

\author{X. Marti}
\affiliation{Institute of Physics ASCR, Cukrovarnick\'a 10, CZ-162 53 Praha 6, Czech Republic}

\date{\today}

\begin{abstract}

Ab initio total energy calculations show that the antiferromagnetic
(111) order is not the ground state for the ideal CuMnSb Heusler alloy
in contrast to the results of neutron diffraction experiments.
It is known, that Heusler alloys usually contain various defects
depending on the sample preparation.
We have therefore investigated magnetic phases of CuMnSb  assuming the most common defects which exist in real experimental conditions.
The full-potential supercell approach and a Heisenberg model
approach using the coherent potential approximation are adopted.
The results of the total energy supercell calculations indicate that defects
that bring Mn atoms close together promote the antiferromagnetic (111) structure already
for a low critical defect concentrations ($\approx$ 3\%).
A detailed study of exchange interactions between Mn-moments further
supports the above stabilization mechanism.
Finally, the stability  of the antiferromagnetic (111) order is enhanced by inclusion of
electron correlations in narrow Mn-bands.
The present refinement structure analysis of neutron scattering experiment
 supports theoretical conclusions.
\end{abstract}


\maketitle

\section{Introduction}

Ferromagnetic Heusler and semi-Heusler alloys are materials with
interesting physical properties.
We refer the reader to a comprehensive recent review.\cite{ha-rev}
Several of them, due to their halfmetallic character, structural
similarity with semiconductors, Curie temperatures above the
room temperature, full spin polarization at the Fermi energy,
and low magnetization damping are materials with potential
in technological applications.
The typical examples of such halfmetallic alloys are
the semi-Heusler NiMnSb alloys, the full Heusler alloy Co$_{2}$FeGa
or Mn$_{2}$CoAl with the inverse Heusler structure.

Halfmetallic Heusler alloys are usually ferromagnets, but there are also
antiferromagnetic (AFM) alloys which can be potential materials
for the so-called antiferromagnetic spintronics.\cite{afm-spin}
Recently the anisotropic magnetoresistance of the AFM-FeRh
alloy was studied both experimentally and theoretically.\cite{afm-ferh}
The antiferromagnetism of CuMnSb Heusler alloys is well established,
\cite{afm-cumnsb} but its low critical N\'eel temperature limits
practical applications as contrasted with the above-mentioned AFM FeRh
alloys.
A new family of CuMnX alloys, where X are elements of the fifth group
with potentially higher critical temperature, was proposed recently
and studied theoretically.
A promising candidate seems to be the layered CuMnAs alloy
with a predicted N\'eel temperature in the range of room temperatures.\cite{cumnas-th,cumnas}

The halfmetallic CuMnSb, the first AFM Mn-based member of Heusler
and semi-Heusler alloy family, therefore attracted the theoretical
interest in the past, in particular as concerns its magnetic structure.
A pilot first-principles study of its magnetic properties was
the subject of Ref.~\onlinecite{cumnsb-th}.
Authors of this paper have compared the total energies of the
non-magnetic, ferromagnetic (FM), and the AFM phases of CuMnSb
at the experimental volume.
They found, in agreement with the experiment, that the ground state is
the AFM one.
The total energy of the FM phase is higher while the total energy of
the non-magnetic phase was found to be significantly above energies
of the AFM and FM phases.
The AFM phase was specifically chosen to be the AFM111 one, with the
alternating (111)-planes of Mn atoms with aligned spins, as known
from experimental neutron scattering studies. \cite{afm-cumnsb}

The fact that the theoretical situation can be more complex was
indicated earlier in the study of (Cu,Ni)MnSb random alloys.
\cite{cunimnsb}
The emphasis there was put on the explanation of unusual concentration
dependence of magnetic moments and residual resistivities, namely, the
dramatic change of the slope of the above concentration dependencies for
the Cu-content of 70-75\%.
However, different ab initio calculations found that the total energy of the ideal
CuMnSb AFM100 structure is lower than the  AFM111 phase detected experimentally.
This result, obtained by the tight-binding linear muffin-tin orbital
(TB-LMTO) method\cite{cunimnsb} was also confirmed by our full-potential
calculations.

Defects are very common in Heusler alloys and their
type and amount depend strongly on the sample preparation and
annealing procedure.
Defects either preserve (swaps)
stoichiometric composition or violate it (impurities).
In the case of CuMnSb alloys, real experimental samples for which the
magnetic structure was studied show relatively large residual
resistivities of order 50 $\mu\Omega$cm (see, e.g., reference in
Ref.~\onlinecite{cumnsb-th}), which clearly indicates the
presence of a non-negligible sample disorder.
It is thus interesting to investigate how the presence of impurities
will influence the magnetic stability of CuMnSb.
A systematic study of formation energies of possible stoichiometric
(swaps) and non-stoichiometric (antisite) defects in a closely related
NiMnSb alloy was the subject of Ref.~\onlinecite{nimnsb-fe}.
We will adapt this approach to the case of CuMnSb to identify
the most probable defects according to their formation energies.
A systematic estimate of formation energies of all possible
defects is, however, beyond the scope of this paper.
We limit ourselves to the most probable ones discussed in experiments
\cite{cumnsb-str} or to those chosen by the analogy with the NiMnSb case
\cite{nimnsb-fe} and perform calculations in the framework of the
TB-LMTO approach combined with the coherent potential approximation (CPA).
The most probable defects seem to be Mn atoms on the
Cu-sublattice, Mn$_{\rm Cu}$, Cu atoms on the Mn-sublattice,
Cu$_{\rm Mn}$, Mn atoms at interstitial sites in the non-stoichiometric
case, and the Cu$\leftrightarrow$Mn swaps for the stoichiometric CuMnSb.

We study the effect of disorder on the magnetic structure due to
the defects by two independent approaches:
(i) the supercell approach employing the full-potential
Vienna ab-initio simulation package (VASP) \cite{vasp}, and
(ii) the Heisenberg model whose parameters were determined
by the TB-LMTO-CPA approach \cite{book} in which the effect of
randomness is included in the framework of CPA.
A main problem with supercell approach is the necessity to consider
all probable AFM phases in the search for the ground
state and also the large number of possible configurations which
should be considered to simulate defect concentration dependencies.
The Heisenberg model which describes directly the relevant magnetic
part of total energy and does not assume any specific magnetic order
as it starts from a completely random moment arrangement, can
roughly indicate possible candidates.
In this way we have extended the search also to the AFM40 phase (the
A$_2$B$_2$-type).\cite{zunger}
The final search is then done using the supercell VASP approach due
to its accuracy.

Another reason for the parallel study using the supercell
and Heisenberg model is the fact that both methods represent
limiting models of disordered systems.
While the CPA provides a good description of completely
disordered alloys neglecting the local environment
or short-range order effects.
The supercell approach, on the contrary, neglects the randomness
of defects.
While both approaches should agree reasonably well in the case
of defects of one type (Mn$_{\rm Cu}$-antisites or Mn-interstitials),
in the case of two defects (Cu$\leftrightarrow$Mn swaps) both models are very different.
There is no correlation between positions of Mn$_{\rm Cu}$- and
Cu$_{\rm Mn}$-antisites in the CPA while in the supercell approach
we assume their fully correlated positions in the whole
crystal (practically the nearest-neighbor positions to have
computationally feasible sizes of supercells).
Clearly, one thus can expect largest differences between
both models for this case.
The supercell approach also allows to check the robustness of
results with respect to possible lattice relaxations which are
neglected in the CPA.
The real alloy is thus somewhere between these two limits.

The important characteristics of any magnetic system are magnetic,
or exchange interactions among magnetic moments which enter the
definition of the Heisenberg model.
In the present case, these interactions are predominantly among
Mn-moments.
They are constructed using the first-principles Lichtenstein mapping
procedure \cite{lie} adapted to the TB-LMTO-CPA approach. \cite{eirev}
It should be noted that this procedure allows to estimate
directly interactions among moments at any pair of sites, and, in
particular, also as a function of varying defect concentrations.
This is very difficult in the supercell case.
It is important to note that in the present case the exchange
integrals are determined from the reference disordered local moment
 (DLM) \cite{dlm} state which describes material above the critical
temperature (the paramagnetic state): no specific magnetic
configuration is thus used in their construction, as emphasized
above.

The most important result of the present study is the finding that
already a low concentration of magnetic defects is able to change
the system total energy in favor of the AFM111 phase so that defects
can stabilize the AFM111 ground state of CuMnSb.
Finally, we have also found that a proper treatment of electron
correlations in narrow Mn $d$-bands \cite{cumnsb-th} is important
and also helps to stabilize the AFM111 phase.

\section{Formalism}
\label{Form}

The ideal CuMnSb has a $C_{\rm 1b}$ crystal
structure which can be represented by four fcc-type sublattices
Cu-Mn-ES-Sb shifted along the [111]-direction of the parent
fcc-lattice consecutively by the distance $\sqrt{3}/4 a$, where $a$ is
the lattice constant and the symbol ES denotes the empty sublattice
(interstitial sites).
The AFM111/AFM100 magnetic structures consist of alternating
[111]/[100] planes of Mn atoms with opposite spins which have
doubled unit size as compared to the original $C_{\rm 1b}$
lattice.
Finally, the AFM40 magnetic structure is a tetragonal structure with
alternating double layers of opposite spins along
the [210]-direction (A$_{2}$B$_{2}$-lattice). \cite{zunger}
In our calculations we use the experimental lattice constant
$a= 6.088$~\AA~\cite{afm-cumnsb} and neglect small volume
changes due to defects.

Two different electronic structure methods were used to
determine the total energy of CuMnSb, both ideal and with
specific defects, namely, the supercell VASP and TB-LMTO-CPA
methods.
Their technical details are summarized below.

The supercell VASP calculations were performed using the projector
augmented wave scheme\cite{paw} with different exchange
correlation potentials.
Only one defect in the unit cell is used for an antisite Mn or Cu atom or for Mn in interstitial position,
while two nearest neighbor Mn- and Cu-antisite defects are used to describe
the Cu$\leftrightarrow$Mn swap.
The magnetic supercell containing 12 atoms simulates an ideal crystal
while supercells containing 24, 48, 96 and for some defects also 192 atoms simulate
disordered alloys with defect concentrations of 12.5\%,
6.25\%, 3.13\% and 1.56\%, respectively.
It should be noted that the presence of defects leads to a small
finite total magnetic moment as contrasted with the exactly zero magnetic moment in the
CPA method.

The Brillouin zone was sampled by $12 \times 12 \times 6$ k-points for
the tetragonal unit cell with 24 atoms, and grids with correspondingly
smaller number of sampling points were used for larger supercells.
Identical large unit cell was used for AFM100, AFM111 and AFM40 magnetic
structures for every defect concentration.
The total energy error was 0.05 meV (per supercell),
the difference between input and output charge densities in the final
iteration of selfconsistent iterations was better than
0.1 me bohr$^{-3}$. For selected configurations we also optimized
the local geometry by atomic force minimization. We found that local
distortions have only weak effect on the total energy differences for
small defect concentrations, therefore the most presented results are
shown for unrelaxed structures. No lattice constant changes due to defects are considered.

In the TB-LMTO-CPA approach the effects of substitutional as well
as magnetic disorders are described by the CPA neglecting
possible lattice relaxations, local environment and clustering
(short range order) effects. \cite{book}
In the disordered case we have used a simple screened impurity model
\cite{sim} to improve the treatment of alloy electrostatics.
The $spdf$-basis and Vosko-Wilk-Nusair exchange correlation potential
\cite{VWN} were used while integration over the full Brillouin zone
was done using about 2000 k-vectors.
The same Wigner-Seitz radii were used for all constituent atoms
including the empty spheres that fulfill interstitial sites.
Calculated total energies in the DLM state were used to estimate
the formation energies of possible defects, swaps for the
stoichiometric CuMnSb and impurities for the non-stoichiometric one.
The selfconsistent DLM potentials were also used to estimate exchange
interactions and to construct the Heisenberg Hamiltonian.

In both approaches we have optionally included the effect of electron correlations
in narrow Mn-subbands in the framework of the LDA+U method. \cite{ldau}
We employ the fully localized limit and make replacement $U_{\rm eff} = U-J$,
where $U$ and $J$ are, respectively, Coulomb repulsion and exchange constants.
We used $U_{\rm eff}= 0.13$ Ry which is similar to the value obtained in
Ref.~\onlinecite{ueff} for a closely related NiMnSb Heusler alloy using
the constrained random-phase approximation.
The magnetic part of total energy is estimated from the
classical Heisenberg Hamiltonian \cite{lie,eirev}
\begin{equation}
H = - \sum_{i,j} \, J_{ij} \,
{\bf e}_{i} \cdot {\bf e}_{j} \ ,
\label{hh1}
\end{equation}
where $i,j$ are site indices, ${\bf e}_{i}$ is the unit vector
pointing along the direction of the local magnetic moment at
site $i$, and $J_{ij}$ is the exchange integral between
sites $i$ and $j$.
The exchange integrals, by construction, contain the atom magnetic
moments, their positive (negative) values being indicative of
ferromagnetic (antiferromagnetic) coupling.
It should be noted that for an evaluation of the exchange
integrals we assume the DLM reference state in which possible
small local moments on Cu, Sb, or interstitial sites vanish.
The only non-zero exchange integrals are thus among Mn atoms.
The DLM state assumes no prescribed magnetic order in the alloy, just
the existence of local magnetic moments on Mn-atoms and conclusions
drawn from it are thus very general.
The DLM state can be naturally implemented in the framework
of the TB-LMTO-CPA. \cite{dlm}

In the case of random alloys the exchange interactions are usually
approximated by the expression
\begin{equation}
\label{hh2}
J_{ij} = \sum_{QQ'} J_{ij}^{QQ'} \eta_{i}^Q \eta_{j}^{Q'}
\end{equation}
in which $\eta_{i}^Q$ are occupation indices ($\eta_{i}^Q = 1$
if site $i$ is occupied by the atom of type $Q$ and $\eta_{i}^Q = 0$
otherwise).

In the general case one should consider the dependence of vectors
${\bf e}_{i}$ on the type of atom at the site $i$.
The situation is simpler in the present case, because on each sublattice
there is at most one type of magnetic atom.
Consequently, the configurationally averaged energy is given by
\begin{equation}
\label{hh3}
\langle H \rangle = - \sum_{i,j} \, \sum_{QQ'} \, J_{ij}^{QQ'} \,
        \langle  \eta_{i}^Q \eta_{j}^{Q'} \rangle \,
        {\bf e}_{i} \cdot {\bf e}_{j} \, .
\end{equation}
Note that $J_{ij}^{QQ'} = 0$ for $i = j$ and the occupation indices
belonging to different sites $i$ and $j$ average independently,
$\langle \eta_{i}^Q \eta_{j}^{Q'} \rangle = x_{i}^Q x_{j}^{Q'}$, where
$x_{i}^Q$ are (local) concentrations.
The averaged total energy $\varepsilon$ per elementary cell, e.g.,
that corresponding to the specific magnetic phase, is
then given by
\begin{equation}
\label{hh4}
\varepsilon  = - \sum_{i}^{\rm cell} \, \sum_{j} \, \widetilde{J}_{ij} \,
        {\bf e}_{i} \cdot {\bf e}_{j} \, ,
\end{equation}
where
\begin{equation}
\label{hh5}
\widetilde{J}_{ij} = \sum_{QQ'} \, J_{ij}^{QQ'}  \, x_{i}^Q x_{j}^{Q'} \,
\end{equation}
are effective exchange interactions.
Calculated energies can be compared with those obtained from the
corresponding total energy calculations based on the supercell VASP
approach.

In the present approach we assume that magnetic moments have
fixed size but allow their different spin directions when searching
for the magnetic ground state.
It should be noted that there exists a more general approach, the magnetic
cluster expansion method \cite{mce}, in which also the size of moments
is a variational parameter.
On the other hand, for an alloy with very rigid magnetic moments like,
e.g., CuMnSb with Mn-moments, the present simpler approach is reasonable
unless higher-order magnetic interactions become important.

\section{ Results and discussion}

\subsection{ Magnetic ground state: the ideal CuMnSb}
\label{MGS_i}

We shall start our discussion by investigating possible magnetic
phases based on the Heisenberg model.
The easiest way of doing so is to calculate the lattice Fourier
transform $J({q})$ of exchange integrals (\ref{hh2}) which enter the
definition of the corresponding Heisenberg Hamiltonian (\ref{hh1}).
 The $J({q})$ describes formally possible spin spirals on a
given lattice.
We shall employ two models: (i) in the first one leading
interactions are obtained from the DLM phase which assumes no
magnetic order, and (ii) two effective interactions are estimated
from total energies of the FM, AFM100, and AFM111 phases calculated
in the VASP.
These interactions represent a fit of the three above energies onto
the Heisenberg model and in some sense they include all interactions
as determined above by first-principles mapping from the DLM phase.
In both cases we used the LDA+U approach.
The values are: $J_{1}/J_{2}$= $-0.30/+0.013$~mRy for the DLM case,
and $J_{1}/J_{2}$= $-0.27/-0.034$~mRy for the VASP.

\begin{figure}[h!]
\center \includegraphics[width=7cm]{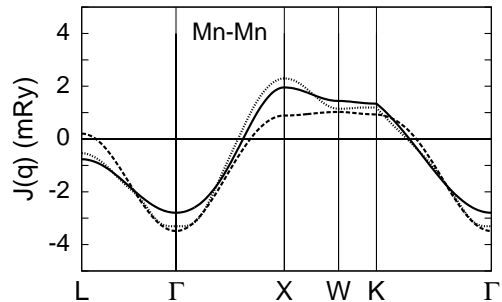}
\caption {Lattice Fourier transform $J(q)$ of the first two
exchange interactions $J_{ij}^{\rm Mn,Mn}$ for the ideal CuMnSb,
obtained for the reference DLM state (full line) and derived from
total energies for the FM, AFM100, and AFM111 phases in the VASP
(dashed line). The case of 62
exchange interactions for the DLM state is shown in dots.
}
\label{fig1}
\end{figure}

The result is shown in Fig.~\ref{fig1} in which, due to the minus
sign in (\ref{hh1}), the higher values correspond to lower
energies and vice versa.
Energies of such spin spirals correspond to specific magnetic
phases, e.g., the values at $q$ = $\Gamma$, X, L, and W relate
to the FM, AFM100, AFM111, and AFM40 phases, respectively.
A general magnetic phase diagram of Heusler alloys in the
$J_{1}/J_{2}$-plane was determined in Ref.~\onlinecite{HHmag}.
 In agreement with its predictions, we have the AFM100 ground
state for the DLM case, but the AMF40 ground state for the VASP, although
the AFM100 phase has almost identical energy.
Both models thus predict that the AFM111 phase is not a ground
state for the ideal CuMnSb.
They also suggest that the AFM111, AFM100, and AFM40 are good
candidates for the ground state (the FM phase can be excluded).
The use of increased number of exchange integrals (62 shells)
in the evaluation of $J(q)$ in the DLM case has only small effect.
Calculations also predict that even more complex spin states
could exist.

\begin{table}[h]
\caption{Total energy in meV per Mn atom for different phases of
ideal CuMnSb and different type of electron exchange and correlation.
The ordered structure AFM111 has been taken as reference $E_{\rm tot}$=0.  \\}
\renewcommand\arraystretch{1.2}\begin{tabular}{lccc}
\hline
 & LDA~\cite{zca}& GGA~\cite{GGA}& LDA+U~\cite{ldau}\\\hline
{AFM100}&$-0.89$&$-0.86$&$-0.52$\\
{ AFM40}&$-1.14$&$-1.03$&$-0.55$\\\hline
\end{tabular}\renewcommand\arraystretch{1.}
\label{t1}
\end{table}

Using the above predictions, we present in Table~\ref{t1} the
total energies of such phases calculated using the VASP method.
While all values shown correspond to the experimental lattice
constant, we have also calculated total energies corresponding
to the volume minimization for each phase and found only small
differences which do not change the result.
The VASP calculations agree well with predictions of
the Heisenberg model.
We mention in particular a correct relation of the AFM40 and AFM100 phases
despite the fact that the former phase was not included in the estimate
of $J_{1}$ and $J_{2}$ interactions.

\subsection{ Formation energies of defects}
\label{FM}

In this section we estimate formation energies of the most
probable defects in CuMnSb, both in the stoichiometric
(swapping defects) and non-stoichiometric (antisite impurities) alloys.
We employ the approach of Ref.~\onlinecite{nimnsb-fe} used for
a related NiMnSb Heusler alloy.
The NiMnSb semi-Heusler alloys have high Curie temperatures
around 750~K and the use of the reference FM state is well
justified.
On the other hand, CuMnSb has a very low (N\'eel) critical temperature
of about 55~K, and in this case a reasonable reference state
is the paramagnetic DLM state.

We will consider the following defects: (i) Mn-antisites on Cu,
Mn-interstitials, i.e., Mn-rich CuMnSb sample, (ii) Cu-antisites
on Mn, i.e., Cu-rich CuMnSb sample, and (iii) the following five
swaps, namely, Cu$\leftrightarrow$Mn, Cu$\leftrightarrow$Vacancy,
Cu$\leftrightarrow$Sb, Mn$\leftrightarrow$Vacancy, and
Mn$\leftrightarrow$Sb.
These defects are considered as dominant ones both in the
experiment \cite{cumnsb-str} and in the related theoretical
study of NiMnSb \cite{nimnsb-fe} (if one interchanges Ni and Cu).
We determine the swap (sw) and defect (def) formation energies (FE) per
formula unit using expressions
\begin{equation}
\label{fesw}
{\rm FE^{\rm sw}} = (E_{\rm tot}^{\rm sw} - E_{\rm tot}^{\rm id})/x_{\rm sw}
\end{equation}
and
\begin{equation}
\label{fedef}
{\rm FE^{\rm def}} = (E_{\rm tot}^{\rm def} - E_{\rm tot}^{\rm id}-
             x_{\rm Cu} \mu_{\rm Cu} - x_{\rm Mn} \mu_{\rm Mn}
             - x_{\rm Sb} \mu_{\rm Sb})/x_{\rm def}  \, ,
\end{equation}
respectively.
Here, $x_{\rm sw}$ is the swap concentration, $x_{\rm def}$ is the concentration
of the defect of the type D, $\mu_{\rm D}$ is the chemical potential of the corresponding
reservoir (D= Cu, Mn, or Sb), and corresponding total energies of ideal CuMnSb,
CuMnSb with swaps, and CuMnSb with defects are denoted as $E_{\rm tot}^{\rm id}$,
$E_{\rm tot}^{\rm sw}$, and $E_{\rm tot}^{\rm def}$, respectively.
It should be noted that contrary to FE$^{\rm sw}$, FE$^{\rm def}$ depend on the
choice of chemical potentials.
We calculate the chemical potentials as the energy per atom of
the reservoir phase, i.e., the fcc-Cu, fcc-DLM Mn (LDA+U), and
Sb in the diamond structure.

\begin{table}[h]
\caption{Formation energies (\ref{fesw}) and (\ref{fedef})
of selected defects (eV per formula unit) in the DLM-CuMnSb
estimated using the TB-LMTO-CPA-LDA+U approach.
ES stands for the vacancy.
Calculations were done for the defect concentration $x=0.05$ in each case.
\\}
\renewcommand\arraystretch{1.2}\begin{tabular}{cccc}
\hline
 Mn$_{\rm Cu}$ & Cu$_{\rm Mn}$ & Mn$_{\rm ES}$ &  Cu$\leftrightarrow$Mn \\
 0.292 & 0.666  & 0.730 & 1.016 \\ \hline
% \multicolumn{4}{c}{ }\\\hline
 Cu$\leftrightarrow$ES & Mn$\leftrightarrow$ES & Mn$\leftrightarrow$Sb & Cu$\leftrightarrow$Sb \\
 2.482  & 3.231  & 3.280 & 7.446 \\ \hline
\end{tabular}\renewcommand\arraystretch{1.}
\label{t2}
\end{table}

In Table~\ref{t2} we show formation energies of three
typical defects for Mn- and Cu-rich CuMnSb alloys.
The most energetically favorable defect is the Mn-antisite on the
Cu sublattice, the Cu-antisite on the Mn sublattice and the
Mn interstitial defect.
The estimated formation energies for some of possible swaps are also shown.
The lowest formation energy corresponds to the Cu$\leftrightarrow$Mn swap similarly
as in the related NiMnSb alloy for Ni$\leftrightarrow$Mn swap.\cite{nimnsb-fe}
It should be noted that present calculations neglect the local
environment effects which, as demonstrated for NiMnSb, generally
lower formation energies as compared to those obtained using
the CPA. \cite{nimnsb-fe}

\subsection{ Magnetic ground state: CuMnSb with defects}
\label{MGS_d}

Non-stoichiometry and intermixing of Cu and Mn can be present in real materials.
Mn atoms become the bonding neighbors of regular Mn atoms with
much stronger exchange coupling that may overbalance the exchange
energy due to interactions within the Mn sublattice, and reverse the
order of total energies for AF100 and AF111. The defects which are present in real samples can stabilize the
 AFM111 phase.
We consider only defects that may appear with the highest
probability, i.e., those with the lowest formation energies.
To this end we employ total energies calculated as functions of
the impurity concentrations for various magnetic states.
The main results were obtained in the framework of the supercell VASP
approach, but we also present results obtained using the Heisenberg
Hamiltonian based on the paramagnetic (DLM) state.
As explained above, the main reason for this is to illustrate the
effect of different treatments of disorder, namely, the supercell
approach and the CPA, on the magnetic stability.
We consider the following defects:
(i) Mn antisites on Cu sublattice,
(ii) Mn interstitials,% on the empty sublattice,
(iii) Cu antisites on Mn sublattice, and
(iv) Mn$\leftrightarrow$Cu swaps.
In the case (iv) the chemical composition is unchanged, while in
the former cases the system is off-stoichiometric, either Mn-rich
(cases (i) and (ii)), or Cu-rich (case (iii)) ones.

\begin{figure}
\begin{center} \includegraphics[width=8cm,angle=0]{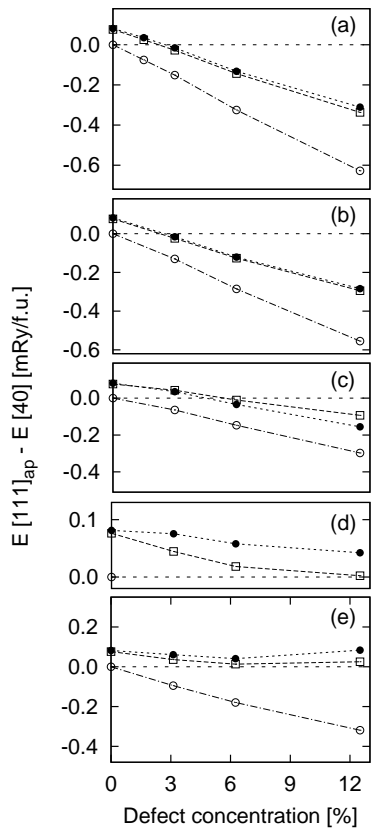}\end{center}
\caption {
The total energy differences $E[111]_{\rm ap} - E[X]$ between
the reference AFM111$_{\rm ap}$ state and corresponding antiferromagnetic
configuration X as a function of concentrations of the following defects:
(a) Mn antisites on Cu; (b) Mn antisites on Cu but with relaxed
 geometry; (c) Mn interstitials; (d) Cu antisites on Mn; and
(e) Cu$\leftrightarrow$Mn swaps as obtained from VASP calculations.
Index ${\rm ap}$ denotes the AFM phase with antiparallel alignment of defect Mn-moments to the moments on the native
Mn-sublattice.}
\label{msvasp}
\end{figure}

\begin{table}[h]
\caption{Ground states of various defects for $x > x_c$, $x_c$ is a critical crossover concentration of defects.
The impurity is Mn$\uparrow$ or Cu.
Mn impurity has additionally 4 Sb nearest-neighbors (NN), which is not shown. Index ${\rm ap}$ denotes the
 AFM phase with antiparallel alignment of defect Mn-moments to the moments on the native Mn-sublattice.
\\}
\renewcommand\arraystretch{1.2}\begin{tabular}{lcc}
defect & ground state & NN atoms of Mn$\uparrow$\\\hline
 Cu$\leftrightarrow$Mn NN swap       & AFM40 & 1 Cu 1 Mn$\uparrow$ 2 Mn $\downarrow$\\
 Cu$\leftrightarrow$Mn swap & AFM111$_{\rm ap}$& 1 Mn$\uparrow$ 3 Mn$\downarrow$ \\
 Mn$_{\rm Cu}$        & AFM111$_{\rm ap}$& 1 Mn$\uparrow$ 3 Mn$\downarrow$ \\
 Mn interstitial  & AFM111$_{\rm ap}$& 1 Mn$\uparrow$ 3 Mn$\downarrow$ \\
 Cu$_{\rm Mn}$        & AFM40& 4 Cu \\\hline
\end{tabular}\renewcommand\arraystretch{1.}
\label{t3}
\end{table}

\begin{figure}
\begin{center}
\includegraphics[width=5.5cm,angle=270]{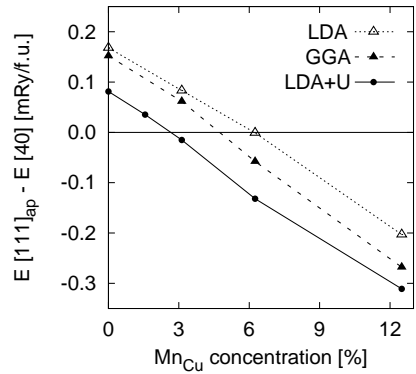}
\end{center}
\caption {The total energy differences of antiferromagnetic configurations
$E [111]_{\rm ap} - E [40]$ calculated with different exchange
and correlation potentials as functions of Mn$_{\rm Cu}$ concentrations.
The results for LDA~\cite{zca}, GGA~\cite{GGA}, and LDA+U~\cite{ldau} are
presented for unrelaxed structures. Index  ${\rm ap}$ denotes the AFM phase with antiparallel alignment of defect
Mn-moments to the moments on the native Mn-sublattice.
}
\label{lda-u}
\end{figure}

The results of the supercell calculations for various defect
types and varying concentrations are summarized in Fig.~\ref{msvasp}
and Table~\ref{t3}.
The following conclusions can be made:
(i) The most favorable defect stabilizing AFM111 phase is the Mn-antisite on Cu lattice (Fig.~\ref{msvasp}a,b).
 For this case we have also shown the effect of lattice relaxations.
The Mn $\uparrow$ substituting Cu prefers the Mn nearest neighbors
with magnetic moments $\uparrow \downarrow \downarrow \downarrow$,
its magnetic moment (lower than magnetic moments of atoms on the
Mn-sublattice) is oriented in our naming convention {\it antiparallel} (see Table~\ref{t3}).
  The critical concentration for which
the AFM111 is stabilized is less than 3\% and the effect of lattice
relaxations is negligible. We thus neglect their effect also for
other defect types.
(ii) Mn interstitials also stabilize the AFM111 phase although
their effect is not so strong, the AFM111 phase is stabilized above about 6\% (Fig.~\ref{msvasp}c).
(iii) Cu antisites on Mn lattice are unable to stabilize the AFM111
phase at low defect concentrations observed in experiment,
although they also slightly promote it (Fig.~\ref{msvasp}d);
(iv) For the case of Mn-antisites on Cu lattice we demonstrate
the effect of electron correlations by comparing the LDA,
GGA, and LDA+U results for AFM40 to AFM111 crossover (see
Fig.~\ref{lda-u}). The critical concentration is reduced about
two times by correlation effects in comparison with the LDA and
slightly less in comparison with the GGA case.
It should be noted that low concentrations of defects which
stabilize the AFM111 phase are on the border of the accuracy of
experimental methods currently used for structural studies, namely,
the X-ray spectroscopy and the neutron scattering experiment (see
Sec.~\ref{Exp} for more details concerning the experiment).

\begin{figure}
\begin{center}
\includegraphics[width=9cm,angle=0]{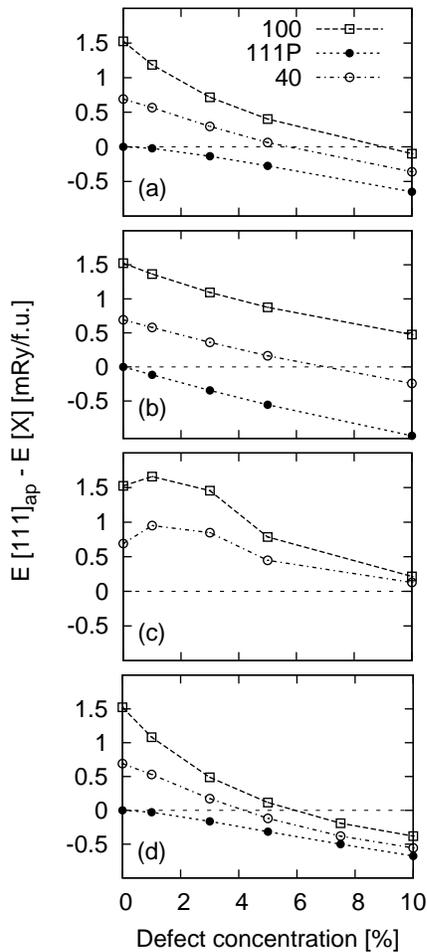}
\end{center}
\caption { The total energy differences $E[111]_{\rm ap} - E[X]$
between the reference AFM111$_{\rm ap}$ state and other antiferromagnetic configurations
for different defect concentrations of Mn antisites on Cu (a), Mn interstitials (b),
Cu antisites on Mn (c) and for Cu$\leftrightarrow$Mn swap (d) as obtained from
Heisenberg Hamiltonian calculations are shown.
Index  ${\rm ap}$ denotes the AFM phase with antiparallel
alignment of defect Mn-moments to the moments on the native Mn-sublattice.
}
\label{nnal}
\end{figure}

Results based on the Heisenberg Hamiltonian are shown in Fig.~\ref{nnal}.
We emphasize that these results were obtained from the DLM reference
state assuming no magnetic preference and neglecting the local environment
effects.
We see a good agreement with the supercell results in particular for
the AFM40 phase for Mn-antisite on Cu, Mn-interstitial, and Cu-antisite
on Mn, although the calculated critical concentrations are larger.

A possible relevance of local environment effects on the magnetic
stability can be illustrated on Cu$\leftrightarrow$Mn swaps (see
Fig.~\ref{msvasp}e and Fig.~\ref{nnal}d for the supercell and
Heisenberg models, respectively).
While the Heisenberg model predicts the stabilization of the AFM111
phase by swaps the supercell approach does not.
The explanation consists in very different ways of modeling of
swaps in both approaches: while in the supercell approach we assume nearest-neighbor
(NN) Cu$\leftrightarrow$Mn swap in the whole crystal,
in the  Heisenberg model based on the CPA we have completely uncorrelated
Cu and Mn antisites.
This result thus shows the dependence of present predictions based on
small unit cell calculations on the local environment effects, in particular
for swaps (see Table~\ref{t3}).
The CPA model is more realistic in this case.
We can summarize that the stabilization of the AFM111 phase by
small concentrations of defects with the lowest formation energies
is an acceptable assumption to explain existing experimental data
confirming the AFM111 state for CuMnSb alloy.
Our model calculations show that the exchange coupling between the NN pairs
of Mn atoms on two different sublattices stabilizes the AFM111 order.
We will therefore analyze exchange interactions in more detail in the next Section.

\subsection{ Exchange interactions}
\label{EI}

Exchange integrals between Mn-moments in both ideal CuMnSb and CuMnSb
with Mn antisites on Cu as a typical example of a possible defect were
estimated using the reference DLM state, the same
as used above for the calculation of total energies of the
Heisenberg model.

We shall start discussion by comparing exchange integrals of ideal
CuMnSb obtained using different reference states, namely, the DLM,
AFM111, and AFM100 states.
The symmetry of the lattice in the AFM100 and AFM111 reference
states is lower as compared to that in the DLM state.
We therefore present their shell-averaged values.
The shell-averaged $J_{ij}^{\rm Mn,Mn}$ are obtained by summing
up all interactions in a given shell with distance $d$ divided by their number.

\begin{figure}
\center \includegraphics[width=6cm]{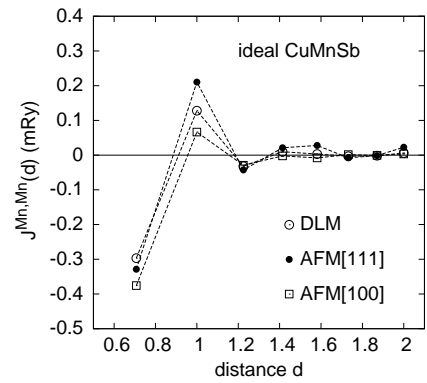}
\caption { Exchange integrals of the ideal CuMnSb as a function
of the distance $d$ (in units of the lattice constant)
between Mn-moments evaluated using three different reference states,
namely, the DLM-state, the AFM111, and AFM100 ones. In the
latter two cases we present shell-averaged values (see text for
details).
}
\label{J_shell}
\end{figure}

The result is shown in Fig.~\ref{J_shell}.
We emphasize that the character of exchange interactions is very
similar in spite of very different reference states.
In particular, we note the negative (AFM-like) first and third
NN interactions and the positive (FM-like) second NN interaction.
The model Heisenberg Hamiltonian phase diagram for Heusler alloys
assuming two or three NN was studied in Ref.~\onlinecite{HHmag}.
A necessary condition for the formation of the AFM111-phase (and
also for the AFM40 one) in the framework of the two NN model is the
negative $J_{2}^{\rm Mn,Mn}$.
On the contrary, the negative $J_{1}^{\rm Mn,Mn}$ and the positive
$J_{2}^{\rm Mn,Mn}$ stabilize the AFM100 phase.
Calculated exchange integrals show, in agreement with the
total energy calculations that the AFM100 phase has a lower total energy
than AFM111 in the ideal CuMnSb alloy.
This conclusion is confirmed for the model with up to 62 NN
integrals used in the analysis of the ground state of the Heisenberg
model.
\begin{figure}
\center \includegraphics[width=6cm]{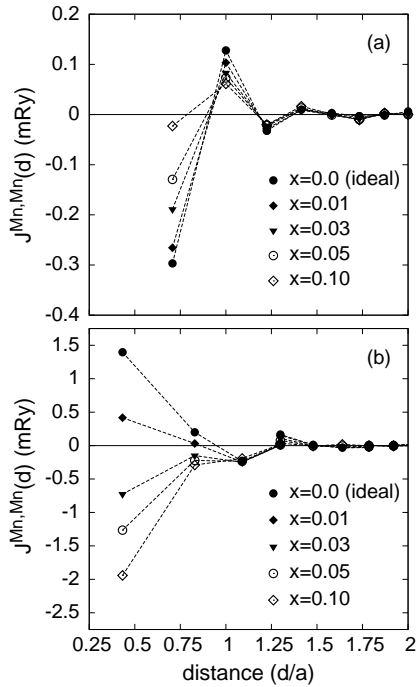}
\caption{Exchange interactions for CuMnSb with varying concentrations
$x$ of Mn antisites on Cu lattice as a function of the distance $d$
between Mn-moments (in the units of the lattice constant) calculated
by using the TB-LMTO-CPA-DLM-LDA+U method:
Interactions among moments on the native Mn-sublattice (a), and
Intersublattice interactions between Mn moments on Mn- and Cu-sublattice (b).
The case $x$=0 corresponds in the
framework of the CPA to the limit of two isolated Mn-moments.
}
\label{fig2}
\end{figure}

There are three kinds of $J_{ij}^{\rm Mn,Mn}$ for CuMnSb with the
nonzero Mn$_{\rm Cu}$-antisite concentrations $x$, namely,
(i) between Mn-moments on the native Mn-sublattice (weight 1, no
defects on the native lattice),
(ii) between Mn-moments on the Cu-sublattice (weight $x^{2}$), and
(iii) intersublattice ones, in which one moment is on the Mn-sublattice
and the other on the Cu-sublattice (weight 2$x$).
Results are shown in Fig.~\ref{fig2} for interactions on the native
Mn-sublattice and for intersublattice ones.
The presence of defects strongly modifies the first two leading
interactions.
It should be noted that $J_{2}^{\rm Mn,Mn}$ is reduced in its size
and becomes comparable to $J_{3}^{\rm Mn,Mn}$ and $J_{4}^{\rm Mn,Mn}$.
This indicates that also more distant interactions
are important.

In particular, the strong weakening of FM-like $J_{2}^{\rm Mn,Mn}$
interactions indicates destabilization of the AFM100 phase with
increasing Mn antisite concentration $x$.
The intersublattice interactions are shown in Fig.~\ref{fig2}b.
The leading intersublattice interaction
is much stronger than on the native lattice
(due to the reduction of NN Mn-Mn distance) and becomes strongly
AFM-like with increasing concentration $x$.
This is also true for the $J_{2}^{\rm Mn,Mn}$ interaction although
less pronounced.
The interactions, whose weight increases with increasing
concentration $x$, are responsible for the crossover % from AFM100
to AFM111 phase.
However, weakening of the positive $J_{2}^{\rm Mn,Mn}$ with
increasing concentration $x$ also supports the crossover.

\subsection{Experiment}
\label{Exp}

CuMnSb samples were prepared by direct synthesis from elements of the
following purity: Cu (99.999\%), Mn (99.98\%) and Sb (99.999\%), mixed
in stoichometric ratio 1:1:1.
Samples were placed in Al$_2$O$_3$ crucibles in double sealed quartz
ampoules under argon atmosphere and heated up to 1000$^{\circ}$~C at a rate
of 1$^{\circ}$~C/min and annealed for one day.
Differential thermal analysis was carried out on Setaram SETSYS-2400 CS
instrument.
The crystal structure of the resulting sample was investigated by means
of X-ray powder diffraction, using Bruker D8 Advance (Cu K$_{\alpha}$
radiation) diffractometer using full-profile Rietveld method.
The X-ray pattern of a CuMnSb sample was indexed in a cubic unit cell
(space group $F\bar{4}3m$ with the lattice parameter $a$ = 6.0974(3) \AA).
The absence of any additional reflections indicates that sample is single phase.
Also the heat capacity measurements were performed in the temperature range
20 - 350 K by means of a Quantum Design PPMS equipment.
For more details see Ref.~\onlinecite{cumnas-th}.

Time-of-flight neutron powder diffraction data were collected at the High Intensity Powder
diffractometer (HIPD) at Los Alamos Neutron Science Center.
Sets of collected diffraction patterns from 6 detector banks
($\pm 40^{\circ}$, $\pm 90^{\circ}$, $\pm 153^{\circ}$) at different temperatures
(10, 20, 30, 40, 50, 60, 70 and 100 K) were used for refinement.
Data analysis was performed using FullProf software\cite{rodr-car}.
Possible magnetic arrangement was determined by irreducible representation
analysis using SARA{\it h} software\cite{wils}.

Structural refinement shows that structure is cubic with space group $F\bar{4}3m$
(216) and cell parameter for 100 K is 6.086(1)~\AA ~as was already reported\cite{endo,afm-cumnsb}.
During the refinement a small discrepancy of intensities of main nuclear
reflections was found.
This can be due to the fact that the atomic occupancies can be different from
the ideal - fully separated - one.
Due to the fact that coherent scattering length of Mn ($b_{c}$ = $-$3.75 fm), Sb ($b_{c}$ = 5.51 fm)
and Cu ($b_c$ = 7.72 fm) are very different, the neutron diffraction data can
provide very accurate information about the mixed site occupancy.
Several possibilities were tested.
The best fit of the experimental data was obtained by mixed occupancies Mn/Cu
(Wyckoff position 4$a$) and Cu/Mn (Wyckoff position $4c$).
Constraints to site full occupancy and same occupancy ratio for both sites
were applied.
By refinement the mix/occupancy ratio was found to be 1.6(1)\%
 in a good agreement with results of presented theoretical calculations.

An apparition of magnetic reflections was observed on diffraction patterns bellow 50 K.
These reflections can be indexed using magnetic propagation vector
 $q$ = ($\frac{1}{2} \frac{1}{2} \frac{1}{2}$).
This means that magnetic unit cell doubles the structural one in all three directions.
From the irreducible representation (IR) analysis only two possible arrangements of
spins for the Mn site (Wyckoff position 4a) are possible - $\Gamma _2$ and $\Gamma _3$.
Both solutions were tested, but only $\Gamma _2$ model was able to fit the
experimental data.
IR $\Gamma_2$, AF111, is represented by (111) ferromagnetic planes with spins perpendicular
to the plane which are stacked antiferromagnetically along [111] direction.
The same configuration can be described by the magnetic space group R$_I$3c
(No. 161.72, setting $\frac{1}{2}a-\frac{1}{2}b$, $\frac{1}{2}b-\frac{1}{2}c$, $2a+2b+2c$).
Calculated magnetic moment of Mn atoms at 10 K is 3.4(1)~$\mu_B$.
This value is very close to 3.9(1)~$\mu_B$ obtained with neutron diffraction at 4 K
and reported in Ref.~\onlinecite{afm-cumnsb}.
The small discrepancy can be due to the different temperatures at which the neutron
diffraction data were collected.

\section{Conclusions}
\label{Con}

The controversy between the experimentally observed AFM111
magnetic structure
of CuMnSb and the total energy calculations which predict
a different ground state for this material is resolved.
Large residual resistivities observed experimentally indicate
presence of structural defects due to the sample preparation.
We suggest as an explanation the presence of defects in real
material.
We have therefore investigated from first principles the ground state
of selected magnetic phases of CuMnSb alloy with defects using the
supercell VASP total energy calculation and the model Heisenberg Hamiltonian
derived from the paramagnetic DLM state with no prescribed magnetic
configuration.
The Heisenberg Hamiltonian, determined from the paramagnetic state,
was used to find possible candidates for the magnetic ground state.
In all cases we used the LDA+U approach to treat approximately the
effect of electron correlations in narrow Mn $d$-bands.
The main conclusions are:
(i) The magnetic ground state of ideal CuMnSb crystal is not the
AFM111 phase;
(ii) The experimentally observed AFM111 phase is the ground state of
CuMnSb samples with Mn defects which occupy the nearest neighbor sites
of the native Mn sublattice, namely, Mn-antisites on Cu lattice, and
Mn-interstitials;
(iii) The crossover to the AFM111 phase is due to the very strong
AFM-like intersublattice interactions among the NN Mn-moments with
distances shorter than those on the native Mn lattice.
The weakening of the ferromagnetic 2nd NN interactions on the native
Mn-lattice due to defects also helps to stabilize the AFM111 phase;
(iv) The supercell approach gives critical concentrations
for the crossover to AFM111 state below 3\% for defects mentioned
above while the Heisenberg model approach predicts the values at least
two times larger; and
(v) Neutron diffraction experiment confirms the existence of low
defect concentrations of about 2\%.
We conclude that the experimentally observed AFM111 phase is stabilized
by defects, in particular those that form the NN pairs with Mn atoms on
the native Mn sublattice (Mn antisites on Cu, Mn interstitials, and possibly
also Cu$\leftrightarrow$Mn swaps).\\

\begin{acknowledgments}
We acknowledge the financial support from the Czech Science Foundation
(Grant No. 14-37427G) and the National Grid Infrastructure MetaCentrum
(project LM2010005) for access to computation facilities.
This work has benefited from the use of HIPD at the
Lujan Center at Los Alamos Neutron Science Center, funded
by the DOE Office of Basic Energy Sciences. Los Alamos
National Laboratory is operated by Los Alamos National Security
LLC under DOE Contract No. DE-AC52-06NA25396.
\end{acknowledgments}


\begin{thebibliography}{99}

\bibitem{ha-rev} T. Graf, C. Felser, S. S. P. Parkin,
Progress in Solid State Chemistry {\bf 39}, 1 (2011).

\bibitem{afm-spin} T. Jungwirth, X. Marti, P. Wadley, and J. Wunderlich,
 Nature Nanotech. {\bf 11}, 231 (2016).

\bibitem{afm-ferh} X. Marti, I. Fina, C. Frontera, J. Liu,
P. Wadley, Q. He, R. J. Paull, J. D. Clarkson, J. Kudrnovsk\'y,
I. Turek, J. Kune\v{s}, D. Yi, J.-H. Chu, C. T. Nelson,
L. You, E. Arenholz, S. Salahuddin, J. Fontcuberta,
T. Jungwirth, and R. Ramesh, Nat. Mater. {\bf 13}, 367 (2014).

\bibitem{afm-cumnsb} R. H. Forster and G. B. Johnston, and D. A. Wheeler,
J. Phys. Chem. Solids {\bf 29}, 855 (1968).

\bibitem{cumnas-th} F. M\'aca, J. Ma\v{s}ek, O. Stelmakhovych,
X. Mart\'i, H. Reichlov\'a, K. Uhl\'i\v{r}ov\'a, P. Beran,
P. Wadley, V. Nov\'ak, and T. Jungwirth,
J. Magn. Magn. Mater. {\bf 324}, 1606 (2012).

\bibitem{cumnas} P. Wadley, V. Nov\'ak, R. Campion, C. Rinaldi,
X. Mart\'i, H. Reichlov\'a, J. \v{Z}elezn\'y, ́ J. Gazquez, M. Roldan,
M. Varela, D. Khalyavin, S. Langridge, D. Kriegner, F. M\'aca,
J. Ma\v{s}ek, R. Bertacco, V. Hol\'y, A. Rushforth, K. Edmonds,
B. Gallagher, C. Foxon, J. Wunderlich, and T. Jungwirth,
Nat. Commun. {\bf 4}, 2322 (2013).

\bibitem{cumnsb-th} T. Jeong, R. Weht, and W. E. Pickett,
Phys. Rev. B {\bf 71}, 184103 (2005).

\bibitem{cunimnsb} J. Kudrnovsk\'y, V. Drchal, I. Turek, and
P. Weinberger, Phys. Rev. B {\bf 78}, 054441 (2008).


\bibitem{nimnsb-fe} B. Alling, S. Shallcross, and I. A. Abrikosov,
Phys. Rev. B {\bf 73}, 064418 (2006).

\bibitem{cumnsb-str} J. Boeuf, C. Pfleiderer, and A. Faisst,
Phys. Rev. B {\bf 74}, 024428 (2006).

\bibitem{vasp} G. Kresse and J. Furthm\"uller, Phys. Rev. B {\bf 54}, 11169
(1996).

\bibitem{book} I. Turek, V. Drchal, J. Kudrnovsk\'y,
M. \v{S}ob, and P. Weinberger, {\it Electronic Structure of
Disordered Alloys, Surfaces and Interfaces}
(Kluwer, Boston, 1997).

\bibitem{zunger} Z. W. Lu, S.-H. Wei, Alex Zunger, S. Frota-Pessoa, and L. G. Ferreira,
 Phys. Rev. B {\rm 44}, 512, (1991).

\bibitem{lie} A. I. Liechtenstein, M. I. Katsnelson, V. P. Antropov,
and V. A. Gubanov, J. Magn. Magn. Mater. {\bf 67}, 65 (1987).

\bibitem{eirev} I. Turek, J. Kudrnovsk\'y, V. Drchal,  and P. Bruno,
Philos. Mag. {\bf 86}, 1713 (2006).

\bibitem{dlm} B. L. Gyorffy, A. J. Pindor, J. Staunton, G. M. Stocks,
and H. Winter, J. Phys. F: Metal Phys. {\bf 15}, 1337 (1985).

\bibitem{paw} G. Kresse and D. Joubert, Phys. Rev. B {\bf 59}, 1758
(1999).

\bibitem{sim} P. A. Korzhavyi, A. V. Ruban, I. A. Abrikosov, and
H. L. Skriver, Phys. Rev. B {\bf 51}, 5773 (1995).

\bibitem{VWN} S. H. Vosko, L. Wilk, and M. Nusair,
Can. J. Phys. {\bf 58}, 1200 (1980).

\bibitem{ldau} A. B. Shick, A. I. Liechtenstein, and W. E. Pickett,
Phys. Rev. B {\bf 60}, 10 763 (1999).

\bibitem{ueff} E. Sasioglu, I. Galanakis, C. Friedrich, and
S. Bl\"ugel, Phys. Rev. B {\bf 88}, 134402 (2013).

\bibitem{mce} M. Yu. Lavrentiev, D. Nguyen-Manh, and S. L. Dudarev,
Phys. Rev. B {\bf 81}, 184202 (2010).

\bibitem{HHmag} J. L. Mor\'an-L\'opez, R. Rodr{\'{\i}}guez-Alba,
and F. Aguilera-Granja, J. Magn. Magn. Mater. {\bf 131}, 417 (1994).

\bibitem{zca} J. P. Perdew and A. Zunger, Phys. Rev. B {\bf 23}, 5048 (1981);
              D. M. Ceperley and B. J. Alder, Phys. Rev. Lett.  {\bf 45}, 566 (1980).

\bibitem{GGA} J. P. Perdew, K. Burke, and M. Ernzerhof,
Phys. Rev. Lett. {\bf 78}, 1396 (1997).

\bibitem{rodr-car} J. Rodriguez-Carvajal, Physica B {\bf 192}, 55 (1993).

\bibitem{wils} A. S. Wills, Physica B {\bf 276}, 680 (2000).

\bibitem{endo}  K. Endo, J. Phys. Soc. Japan {\bf 29}, 643 (1970).


\end{thebibliography}
\end{document}